# Continuous Phase Modulation Technology Based on Grating Period Interval for High Grating Coupling Efficiency and Precise Wavelength Control

Yiming Sun[1*], Simeng Zhu[1], Bocheng Yuan[1], Yizhe Fan[1], Mohanad Al-Rubaiee[1], Xiao Sun[1], John H. Marsh[1], Stephen J. Sweeney[1], Lianping Hou[1]

[1]Jams Watt School of Engineering, University of Glasgow, Glasgow G12 8QQ, UK
*Corresponding author: 2465522S@student.gla.ac.uk

**A novel grating modulation technique for laser arrays is proposed and demonstrated. This method modifies the initial phase within each grating period, applying a total phase shift that increments in an arithmetic progression, ensuring equal channel spacing across the array. Despite the varying phase shifts, the device maintains coupling efficiency comparable to traditional uniform grating structures. Furthermore, the continuous phase modulation enhances the stability of the lasing wavelength of the primary mode, reducing sensitivity to fabrication errors. This improved tolerance to manufacturing inaccuracies represents a significant technological advancement, making this approach highly promising for applications requiring precise and stable wavelength control.**

With the rapid growth in global data transmission demands, the need for efficient and high-capacity optical communication systems has become increasingly urgent. Multi-channel transmission technologies, such as Dense Wavelength Division Multiplexing (DWDM), play a pivotal role in addressing this challenge by allowing simultaneous data transmission over multiple, closely spaced wavelength channels [1]. To support such systems, laser arrays that provide precise, stable, and reliable light sources for each wavelength channel are critical [2]. Traditionally, a distributed feedback (DFB) laser array is created by varying each grating period of lasers to achieve the desired wavelength spacing [3]. However, narrow channel spacing laser arrays, such as 0.8 nm, achieving this through direct adjustments to the grating period is limited by the resolution constraints of electron beam lithography (EBL), typically 0.5 nm. This makes it difficult to fabricate laser arrays with closely spaced channels using conventional grating period tuning.

To address this limitation, sampled grating designs based on reconstruction equivalent-chirp (REC) technology have been introduced. These designs employ sample period modulations to achieve precise wavelength selection and control, overcoming the resolution limits of EBL. Although this technique provides improved wavelength precision, its low coupling efficiency hampers overall device performance and leads to an increase in size [3]. Further optimization of grating structures, such as the introduction of multi-phase shift modulation within each sampling period, has improved both wavelength control and coupling efficiency. Nevertheless, the complexity of its design may cause new manufacturing problems due to the reactive ion etching (RIE) hysteresis effect during the sidewall grating fabrication process, which is not conducive to the simultaneous integration of the laser array [4, 5].

In this work, we introduce a novel grating modulation scheme that employs continuous phase modulation based on the grating period. Different from the classic multi-phase shift grating, this method introduces a fixed phase offset at the onset of each grating period in a reference uniform grating, resulting in an arithmetic progression of total phase shifts across adjacent lasers. This phase-continuous modulation not only ensures equal channel spacing but also significantly enhances the stability of the lasing wavelength under high coupling efficiency by preventing the generation of higher-order modes, similar to the performance of a uniform grating. More importantly, the minor change of its wavelength is based on the disturbance of its overall phase, since small discrepancies in the manufacturing process, such as lithography or etching imperfections, have a minimal impact on the device's performance. This makes the proposed method particularly promising for high-performance optical communication systems where wavelength stability and accuracy are essential.

Fig. 1(a) presents the structure schematic of the uniform grating in a single period at the position of 0, $L/4$, $L/2$, $3L/4$ and $L$, where $L$ is the total length of the grating. The uniform grating structure typically consists of a grating region with a 0.5 duty cycle and a corresponding non-grating region, creating periodic refractive index perturbations along the cavity. These perturbations facilitate light coupling and enable lasing. The consistent periodicity of the grating ensures uniformity within each period, allowing the lasing wavelength to be adjusted by modifying the grating period. Although this design is simple and effective, the variation of the grating period is limited by the manufacturing accuracy, which introduces challenges when aiming for narrow channel spacing or fine wavelength tuning in multi-channel laser arrays.

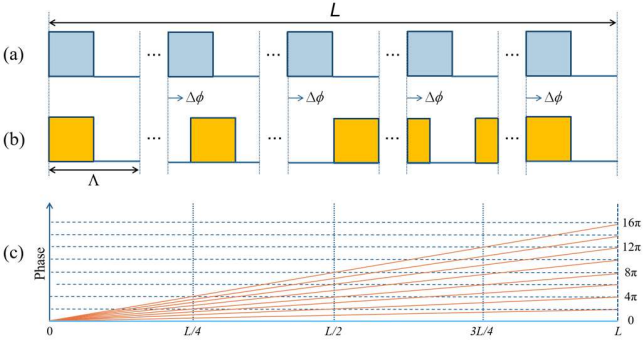

Fig. 1. Schematic of the single-period grating structure at positions 0, $L/4$, $L/2$, $3L/4$ and $L$ for (a) uniform grating and (b) phase modulation within the grating period. (c) The phase increment for the DFB array is $2\pi$, resulting in a phase shift ranging from 0 to $16\pi$ from CH1 to CH8.

As a comparison, Fig. 1(b) illustrates the grating structure utilizes continuous phase modulation within each grating period. Similar to the uniform grating, this design maintains a duty cycle of 0.5, with equal grating and non-grating regions. However, a fixed phase shift $\Delta\phi$ is introduced at the beginning of each grating period, $\Delta\phi$ can be defined as $\Delta\phi = N\Lambda\pi/L$, where $N$ is any real number and $\Lambda$ means grating period. For uniform grating = 0, indicating no phase shift. In Fig. 1(b), $N=2$ indicates that a total phase shift of $2\pi$ is evenly distributed along the entire grating cavity length, $L$. When the grating is positioned at approximately one-quarter of the cavity length ($L/4$), it is centered within the grating period with a phase shift of $\pi/2$. As the phase progresses to around the midpoint of the cavity ($L/2$), the phase shift reaches $\pi$, causing the grating to move toward the right within the period. With continued phase modulation, the grating shifts further to the right in each cycle. Any portion of the grating that exceeds the boundary of the grating period reappears on the left side, ensuring the continuity of the phase modulation. By the time the grating reaches the end of the cavity length ($L$), it has fully shifted back to the left side, completing a total phase shift of $2\pi$. The smooth and continuous nature of the phase shifts within each grating period characterizes the structure as a Continuous Phase Shift Grating (CPSG). The phase distribution, with a total phase shift of $2\pi$ along the CPSG, is illustrated in Fig. 1(c).

The reflection spectrum resulting by continuous phase modulation technique can be accurately calculated using the transfer matrix method (TMM). Fig. 2(a) shows the simulated reflection spectrum of uniform grating and $4\pi$ to $28\pi$-CPSG, with a grating period of 243 nm and a cavity Length of 800 μm. As shown in the figure, the central wavelength position of CPSG increases linearly with the total phase shift increases. The amount by which the wavelength shifts is directly proportional to the magnitude of the phase change, meaning that larger phase shifts result in greater wavelength movement. This linear relationship between phase shift and wavelength position ensures precise control over the lasing wavelength. Compared to a uniform grating, the modulation of CPSG primarily shifts the central wavelength, while other spectral properties, such as bandwidth and reflectivity, remain unchanged. Since the conversion relationship between phase and grating period is $\Delta\phi/\Delta\Lambda = 2\pi/\Lambda$. The wavelength difference between each CPSG can be expressed as:

$$\Delta\lambda = \frac{N \cdot n_{eff} \cdot \Lambda^2}{L} \quad (1)$$

where $n_{eff}$ is the effective refractive index of the ridge waveguide, $N$ is any real number, indicating that the total number of $N\pi$ is evenly distributed across each grating cavity length. For the designed and calculated parameters, $n_{eff}$ is 3.21 and $\Lambda$ is 243 nm placing the lasing wavelength under uniform grating conditions($N=0$) at 1560 nm, which corresponds to the gain peak of the multiple quantum well (MQW) material used. From Eq. (1), the corresponding lasing wavelengths of CPSG for different values of $L$ or $N$ are presented in Table 1. Once the epilayer and grating structure are confirmed, we can accurately obtain $\Delta\lambda$ by adjusting $N$, $\Lambda$, and $L$, even for DWDM wavelength spacing of 0.8 nm or as narrow as 0.4 nm.

**Table 1. The wavelength of CPSG under different conditions**

| Grating Type | Length of Cavity $L$ | Wavelength |
|---|---|---|
| Uniform Grating | 800 μm | 1560.000 nm |
| $2\pi$-CPSG ($N=2$) | 800 μm | 1560.475 nm |
| $4\pi$-CPSG ($N=4$) | 800 μm | 1560.950 nm |
| $4\pi$-CPSG ($N=4$) | 400 μm | 1561.900 nm |

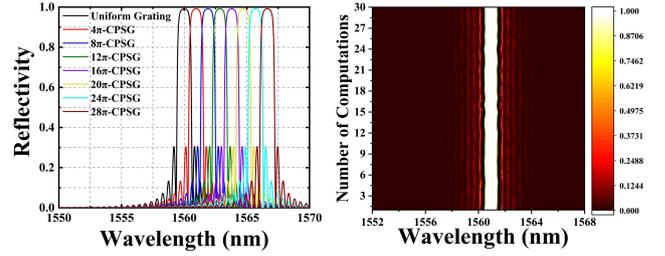

Fig. 2. (a) The reflection spectrum of uniform grating and $4\pi$ to $28\pi$-CPSG, with the same grating period (243 nm) and cavity Length (800 μm). (b) Calculated 2D optical spectra considering the effects of random phase errors.

Different from traditional refractive index modulation, the central wavelength for continuous phase modulation is less dependent on the exact physical positioning of the grating structures. The phase is continuously modulated across the grating cavity length, smoothing out the effects of minor lithographic errors. While the lithographic precision still affects the exact phase shift, the continuity of phase modulation brings a certain degree of compensation. As long as the total phase shift is maintained with a fixed phase interval, minor phase errors between intervals do not affect the central output wavelength of the device. Fig. 2(b) shows the calculated 2D optical spectra with the effect of these tiny random phase errors ($<2N\pi R^2/\Lambda^2$, i.e., maximum phase error per grating period, where $R$ represents the resolution of EBL). Results from 30 repeated calculations show that these small deviations primarily impact the side mode suppression ratio (SMSR), with negligible effect on the central wavelength, which remains stable and unchanged. This demonstrates the robustness of the CPSG design, as the central wavelength is unaffected by phase errors.

To verify the feasibility of the proposed design, laser arrays with varying grating lengths (800 μm and 400 μm) and total phase distributions ($2\pi$ and $4\pi$) were designed and fabricated. In each array, the first laser utilized a uniform grating as a control reference. Subsequent lasers within the array were subjected to phase modulation, the total phase difference between each adjacent laser is $2\pi$ or $4\pi$. The devices were fabricated using an epitaxial structure based on the AlGaInAs/InP material system, which features five quantum wells (QWs) and six quantum barriers (QBs). The detailed epilayer structure is referenced in [6] and the device fabrication process closely follows that described in [7].

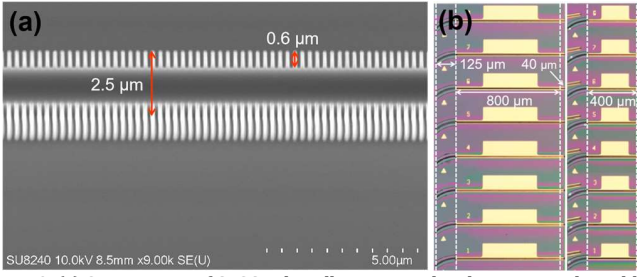

Fig. 3. (a) SEM picture of CPSG sidewall grating with ridge waveguide width of 2.5 μm and a recess depth of 0.6 μm, (b) Optical microscope image of the fabricated DFB laser array featuring different lengths (800 μm and 400 μm).

Figure 3(a) shows a scanning electron microscopy (SEM) picture of a CPSG sidewall grating structure with a ridge waveguide width of 2.5 μm and a grating recess depth of 0.6 μm on each side of the ridge. Due to the small incremental for each phase change, the difference between CPSG and uniform grating structure is difficult to distinguish under the electron microscope. Compared with other phase modulation schemes, this subtlety in the phase modulation also minimizes etching difficulty, and its coupling coefficient is same as that of a uniform grating. Meanwhile, a π phase shift of the seed grating period Λ was inserted into the center of the cavity to ensure single longitudinal mode (SLM) operation of the device [8].

Fig. 3(b) shows an optical microscopy picture of the fabricated DFB laser array. The device is composed of three key sections: a central grating region, a cleave section, and a back end designed to minimize reflection. The central portion features the uniform grating or CPSG with different lengths (800 μm and 400 μm), which serves as the core of the device for wavelength selection and modulation. At the output end, a 40 μm long straight waveguide is included to facilitate clean cleavage after fabrication, ensuring efficient light emission and reducing output scattering. On the opposite side, the back end of the device incorporates a 33° angled curved waveguide with a length of 125μm. This angled design effectively suppresses back reflections, preventing unwanted feedback into the active region and enhancing the overall stability and performance of the laser. In the final stage of fabrication, the sample was cleaved into individual laser bars and was mounted epilayer-up on a copper heat sink on a Peltier cooler. The heat sink temperature was set at 20°C and the devices were tested under CW conditions.

Fig. 4(a) shows an 8-channel laser array with a grating length of 800 μm, where the first laser (CH1) utilizes a uniform grating, and the remaining lasers have a total phase difference of 4π between each consecutive channel. The spectrum was measured using an optical spectrum analyzer (OSA) with a resolution bandwidth of 0.06 nm. With $I_{DFB}$ = 200 mA, the lasing wavelengths are 1564.75 nm, 1565.70 nm, 1566.66 nm, 1567.64 nm, 1568.54 nm, 1569.50 nm, 1570.50 nm and 1571.37 nm from CH1 to CH8.

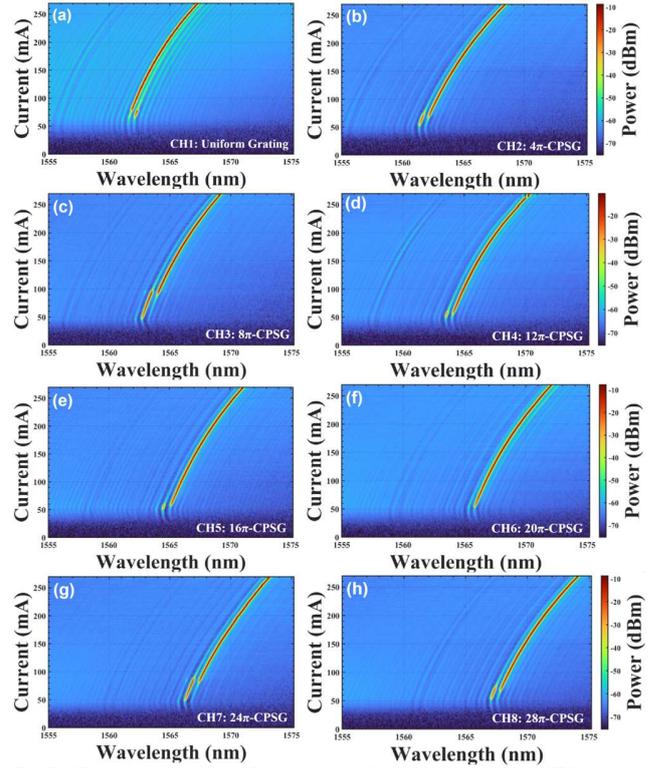

Fig. 5. 2D optical spectra vs $I_{DFB}$ for (a) uniform grating, (b) 4π-CPSG, (c) 8π-CPSG, (d) 12π-CPSG, (e) 16π-CPSG, (f) 20π-CPSG, (g) 24π-CPSG, (h) 28π-CPSG with the grating length of 800 μm and the total phase interval of each adjacent lasers is 4π.

Fig.4(b) shows the corresponding linear fit of the lasing wavelengths of the eight channels, and the slope of the line is 0.949 nm, closely matching the simulated value of 0.95 nm (shown in Table 1), the excellent agreement between experimental and theoretical results shows the strong wavelength accuracy of CPSG design.

Fig. 5 shows a two-dimensional (2D) optical spectrum versus $I_{DFB}$ (0-270 mA) for CH1 to CH8 with a grating cavity length of 800 μm. The threshold current for all devices is approximately 50 mA. At low current levels, side modes exhibit lower threshold conditions, which dominate the lasing behavior initially. Notably, the range of side mode lasing in CPSG may be larger than that of uniform grating (CH1). This may be due to process errors, which are consistent with the characteristics in Figure 2. As the current increases, the gain in the main mode surpasses that of the side modes, leading to the main mode lasing. Very stable SLM operation of laser array is observed from 100 mA to 270 mA, with no mode-hopping. The average current induced wavelength redshift coefficient of all devices is around 0.025 nm/mA, indicating a predictable and consistent current-induced wavelength shift across all channels.

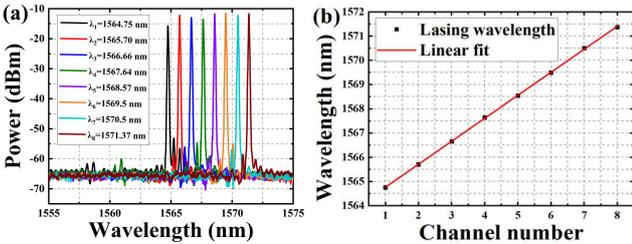

Fig. 4. (a) Measured optical spectrum of the devices for laser array with a total phase difference of 4π between each consecutive channel and a grating cavity length of 800 μm at $I_{DFB}$ = 200 mA, (b) lasing wavelengths along with a linear fit applied to the eight data points.

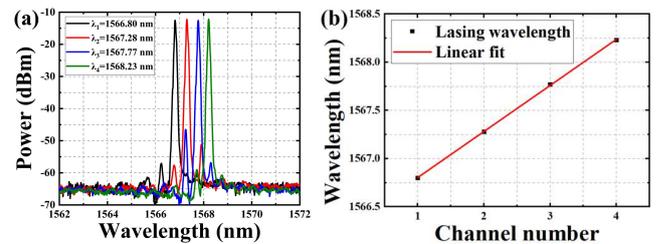

Fig. 6. (a) Measured optical spectrum of the devices for laser array with a total phase difference of 2π between each consecutive channel and a grating cavity length of 800 μm at $I_{DFB}$ = 200 mA, (b) lasing wavelengths along with a linear fit applied to the four data points.

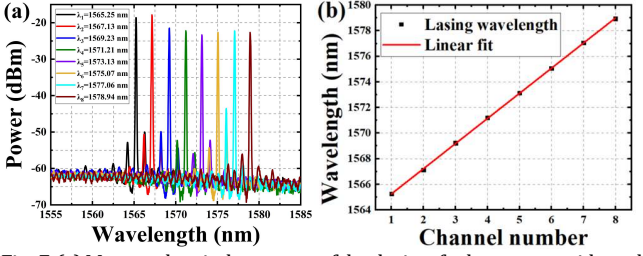

Fig. 7. (a) Measured optical spectrum of the devices for laser array with total phase difference of 4π between each consecutive channel and a grating cavity length of 400 μm at $I_{DFB}$ = 140 mA, (b) lasing wavelengths along with a linear fit applied to the eight data points.

Fig. 6(a) shows the optical spectra for the total phase difference of 2π between each consecutive channel with a grating cavity length of 800 μm. Due to cleavage errors during fabrication, part of the waveguide in the laser array was damaged, limiting the functionality to only 4 out of the 8 lasers operating as expected. With $I_{DFB}$ = 200 mA., the lasing wavelengths are 1566.80 nm, 1567.28 nm, 1567.77 nm and 1568.23 from CH5 to CH8 (corresponding to 8π-CPSG to 14π-CPSG).

Fig. 6(b) shows the corresponding linear fit to the lasing wavelengths of the four channels, and the slope of the line is 0.493 nm. This result confirms the expected wavelength spacing between the channels (0.475 nm), The value of this error (0.02 nm) is less than the resolution of the OSA (0.06 nm), validating the universality of the designed phase shift modulation. Although only half of the array operated as intended, the successful channels demonstrate the feasibility and wavelength tunability of the phase-modulated laser design.

To verify the effect of grating length on the wavelength in equation (1), a laser array was fabricated with a grating length of 400 μm. Similarly, the first laser features a uniform grating, while the remaining lasers have a total phase interval of 4π between consecutive channels. Fig. 7(a) shows the optical spectra for each channel with $I_{DFB}$ = 140 mA. The lasing wavelengths are 1565.25 nm, 1567.13 nm, 1569.23 nm, 1571.21 nm, 1573.13 nm, 1575.07 nm, 1577.06 nm and 1578.94 nm from CH1 to CH8.

Fig.7(b) shows the corresponding linear fit to the lasing wavelengths of the four channels, and the slope of the line is 1.956 nm, with an error of 0.056 nm compared with the designed wavelength spacing of 1.9 nm. This error value is also below the resolution of the OSA, highlighting the excellent wavelength precision achievable with the CPSG structure.

The operation of the short cavity length device demonstrates that the CPSG structure exhibits an effective coupling coefficient (κ) comparable to that of traditional grating designs. To compare the κ of the uniform grating and CPSG structures, the measured spectra with $I_{DFB}$= 50 mA are shown in Fig. 8. From coupled mode theory, the κ can be calculated from $\Delta\lambda_s$ and $\lambda_B$ using the following formula [9]:

$$\kappa = n_{eff} \cdot \frac{\Delta\lambda_s}{\lambda_B^2} \qquad (2)$$

where $\Delta\lambda_s$ is the stop bandwidth; $\lambda_B$ is the Bragg wavelength of the grating. Figures 8(a) and 8(b) show the measured optical spectrum for 400-μm-long DFB lasers with a uniform grating and 4π-CPSG, respectively. For the uniform grating, the stop bandwidth $\Delta\lambda_s$ is 2 nm and the Bragg wavelength $\lambda_B$ is 1562.79 nm. For CPSG, the stop bandwidth $\Delta\lambda_s$ is also 2 nm, and the Bragg wavelength $\lambda_B$ is approximately 1564.09 nm. Therefore, the κ values of the uniform grating and CPSG devices are calculated as 26.29 $cm^{-1}$ and 26.24 $cm^{-1}$, respectively. This demonstrates that the introduction of continuous phase modulation does not compromise coupling efficiency, which remains on par with traditional uniform grating structures. This makes CPSG a highly effective alternative for laser arrays and integrated photonic devices.

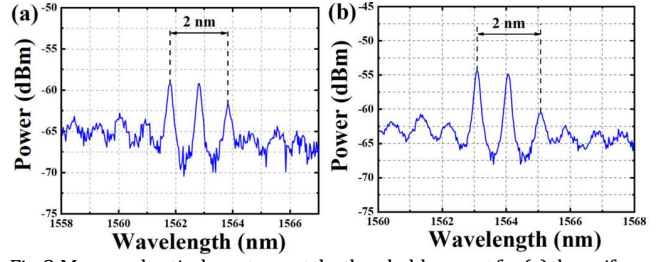

Fig. 8. Measured optical spectrum at the threshold current for (a) the uniform grating and (b) the 4π-CPSG structure, both with a cavity length of 400 μm.

In conclusion, CPSG devices demonstrate significant advantages in terms of stability, robustness, and performance. Despite the introduction of continuous phase modulation, the coupling coefficient of CPSG remains consistent with that of uniform gratings, ensuring efficient optical interaction and light coupling. The phase modulation technique allows for precise wavelength control and reduced sensitivity to fabrication errors, making CPSG devices highly reliable for multi-channel laser arrays. Furthermore, CPSG designs can maintain strong lasing performance even in short-cavity configurations, making them suitable for compact, high-density integrated photonic applications. Overall, the CPSG approach offers a promising solution for advanced optical communication systems and photonic integration, where precision and robustness are paramount.

**Funding.** This work was supported by the U.K. Engineering and Physical Sciences Research Council (EP/R042578/1).

**Acknowledgements.** We would like to acknowledge the staff of the James Watt Nanofabrication Centre at the University of Glasgow for their help in fabricating the devices.

**Disclosures**. The authors declare no conflict of interest.

**References with titles**